\documentstyle[12pt,aasms4]{article}
\begin{document}

\title{ {\it HST} FINE GUIDANCE SENSOR ASTROMETRIC PARALLAXES FOR THREE DWARF NOVAE: SS AURIGAE, SS CYGNI, AND U GEMINORUM}
                                     
\author{Thomas E. Harrison$^{\rm 1}$, Bernard J. McNamara}
\affil{Astronomy Department, New Mexico State University, Las Cruces, NM 88003}
\authoremail{tharriso@nmsu.edu, bmcnamar@nmsu.edu}
                                     
\author{Paula Szkody}
\affil{Department of Astronomy, University of Washington, Seattle, WA 98195}
\authoremail{szkody@alicar.astro.washington.edu}
                                     
\author{Barbara E. McArthur, G. F. Benedict}
\affil{McDonald Observatory, University of Texas at Austin, Austin, TX 78712}
\authoremail{mca@barney,as.utexax.edu,fritz@clyde.as.texas.edu}
                                     
\author{Arnold R. Klemola}
\affil{University of California Observatories/Lick Observatory, University of California, Santa Cruz, California, 95604}
\authoremail{klemola@ucolick.org}
\and                                     
\author{Ronald L. Gilliland}
\affil{Space Telescope Science Institute, 3700 San Martin Drive, Baltimore, MD 21218}
\authoremail{gillil@stsci.edu}

\begin{abstract}
We report astrometric parallaxes for three well known dwarf novae obtained using
the Fine Guidance Sensors on the Hubble Space Telescope. We found a parallax 
for SS Aurigae of $\pi$ = 5.00 $\pm$ 0.64 mas, for SS Cygni we found $\pi$ = 
6.02 $\pm$ 0.46 mas, and for U Geminorum we obtained $\pi$ = 10.37 $\pm$ 0.50 
mas. These represent the first true trigonometric parallaxes of any dwarf novae.
We briefly compare these results with previous distance estimates. This program
demonstrates that with a very modest amount of {\it HST} observing time, the 
Fine Guidance Sensors can deliver parallaxes of unrivaled precision.  
\end{abstract}

\keywords{Astrometry --- novae, cataclysmic variables ---  stars: individual 
(SS Aurigae, U Geminorum, SS Cygni)}

\altaffiltext{1}{Based partially on observations obtained with the Apache Point Observatory 3.5-meter telescope,
which is owned and operated by the Astrophysical Research Consortium.          }

\section{INTRODUCTION}

Like many astronomical objects, the distances to cataclysmic variables are 
imprecisely known. Cataclysmic variables (CVs) are interacting binaries composed
of a white dwarf and a main sequence companion. The secondary star fills its 
Roche Lobe, and matter is transferred to the white dwarf through the inner Lagrangian point. In non-magnetic systems, the mass transfer
occurs through an accretion disk. Most, but not all, CVs exhibit outbursts where the system
suddenly brightens (see Warner 1995 for a complete review of the behavior of the different
subclasses of the CV family). These eruptions range from the luminous classical novae, where
the outbursts are due to a thermonuclear runaway on the white dwarf (see Starrfield et al. 1998),
and the explosion releases E$_{\rm tot}$ $\sim$ 10$^{\rm 45}$ erg over the 
lifetime of the outburst. To the dwarf novae
(DN), where the outburst energy is more modest, E$_{\rm tot}$ $\sim$ 10$^{\rm
40}$, and is generated by an accretion
disk instability cycle (see Cannizzo et al. 1998, and references therein). But our knowledge of
these energies and other fundamental parameters of CV systems suffers due to 
the lack of accurate distance measurements.

Berriman (1987) has compiled a list of distance estimates for CVs that used a 
variety of techniques, including reports of astrometric parallaxes, and 
spectroscopic parallaxes that relied
on the photometric parameters of the CV secondary stars. This latter technique may be the most
reliable, but its application is made uncertain by the complex spectral energy distribution of CVs
at minimum light, where emission from the hot white dwarf, the accretion disk, the irradiated
secondary star, and features that arise from the accretion stream and its impact with the accretion
disk (the ``hot spot'') contaminate the systemic luminosity. To evaluate the accuracy of the various
secondary distance estimators requires direct measurements of the parallaxes of a number of well
known CVs. Of the dwarf novae with astrometric parallaxes compiled by Berriman, only that for
SS Cyg has a significance greater than three sigma: 50 $\pm$ 15 pc (Kamper, 1979). We show below
that even this measurement is incorrect, and we conclude that until now, no dwarf nova has truly
had its trigonometric parallax measured. In this paper we report the first high precision parallaxes
for three well known dwarf novae: SS Aurigae, SS Cygni and U Geminorum. These parallaxes
were obtained using the Fine Guidance Sensors (FGS) on the Hubble Space Telescope ({\it HST}).

\section{FGS ASTROMETRIC OBSERVATIONS}

The FGS were designed to provide exceptional pointing and tracking stability for
the science instruments on the {\it HST} no matter where the telescope was pointed. 
As such, the FGS were designed to have a large dynamic range and a large field-of-view. For normal guiding
operations, only two FGS are employed. This frees the third FGS (``FGS3'') to make astrometric
measurements. There are two modes of astrometric operation: ``position'' and 
``transfer''. For
obtaining parallaxes and proper motions, position mode is used. For resolving close binaries,
transfer mode is used. Astrometry using the FGS has been fully described elsewhere (e.g.,
Benedict et al. 1994, Benedict et al. 1992, Bradley et al. 1991), however, the most comprehensive
source for understanding the nuances of obtaining high precision parallaxes with the FGS is
found in the {\it Fine Guidance Sensor Instrument Handbook}. Since our observing program followed
the prescription outlined within the {\it Handbook}, only the most important details relevant for
parallax measurement will be addressed here. We present a more complete discussion of the FGS
astrometric data analysis for the program CVs, including their proper motions, and comparison
of their astrometric and infrared spectroscopic parallaxes, in Harrison et al. (1999).

As in a classical parallax program, an FGS program consists of measuring the position
of the target object with respect to a number of field stars obtained at several widely separated
epochs when the parallax factors approach unity. Each FGS has a field-of-view (FOV) that
consists of a quarter annulus of inner and outer radii of 10 and 14 arcmin, respectively. This
entire area, known as a ``pickle'', is accessible to the interferometer. Not all of the area of the
pickle ends up being accessible in the typical astrometry project because of the different roll
angles of the {\it HST} found at different epochs of observation. The large FOV remains one of the
greatest strengths of the FGS: It allows astrometry of objects in regions where the density of
potential reference stars is low. It is important to note, however, that the astrometric precision
is a function of the location of the object within the FOV of the FGS. Objects nearer the center
of the pickle are more precisely located than those near an edge. This variation arises from
optical distortion across the FOV, known as the Optical Field Angle Distortion (``OFAD''), which
is greater, and less-well calibrated, near the edges of the pickle (see Jeffreys et al. 1994). A
dedicated calibration program, called the Long Term Stability Test (the ``LSTAB'', see McArthur
et al. 1997), is periodically run throughout each {\it HST} cycle to assess the stability of the OFAD.
The LSTABs detect any changes in plate scale and are vital for astrometric data reduction.

     Another important aspect of the FGS is their dynamic range: Positions of objects with 8.0
$<$ V $<$ 17.0 can be measured without changing the instrument configuration. This is especially
important for the dwarf novae in our program, as at outburst they approach 
V $\sim$ 8, but at
minimum have V $\sim$ 14.5. Thus, we were able to ignore the variability of our targets in planning
the observational program. The astrometric precision to which a target can be measured is a
function of the brightness of the target. For objects with V $<$ 15.0, the single measurement
precision is $\approx$ 1 mas, for fainter objects the precision is $\approx$ 2 mas. The entire error budget for a
minimalist FGS parallax program is $\approx$ 1 mas. A carefully planned program with multiple
observational epochs, like that detailed below, can achieve parallaxes with sub-mas precision.

\section{THE OBSERVATIONAL PROGRAM}

An ideal FGS parallax program would consist of five or more reference stars 
having V $<$ 15.0, all within a few arcminutes of each other, and spread 
uniformly around the target. Such fields are not always available for objects 
of astrophysical interest, and therefore the choice of
potential astrometry targets is limited. We based our selection of CVs on four 
criteria, 1) their astrophysical significance, 2) their minimum brightness (V 
$<$ 15.0), 3) the availability of a good
set of reference stars, and 4) the likelihood that their parallaxes would have a precision of $\leq$ 10\%.
We also confined our selection of CVs to U Gem type DN--objects that have similar outburst
characteristics and where the secondary star is clearly visible at minimum light. This latter
criterion was imposed to allow us to evaluate the accuracy of the technique of spectroscopic
parallax. Clearly, parallaxes for a large sample of CVs would be desirable, but given the scarcity
of {\it HST} time, a modest program was proposed to confirm that high-precision parallaxes for
several such objects could be obtained with a modest amount of observing time.

Three target DN that met the above criteria were SS Aur, U Gem, and SS Cyg. U 
Gem and SS Cyg are very well known, having been observed continuously for more 
than 100 yr. SS Aur is probably less well known outside of the CV community, 
but it is a regularly studied DN (c.f., Shafter and Harkness 1986, Tovmassian 
1987, and Cook 1987). [For further information on these and other DN systems, 
as well as outstanding problems in CV research, the reader is directed to 
Warner (1995).] Using the {\it HST Guide Star Catalogue}, potential reference 
stars 
were identified that fell within the pickle centered on the target DN. We then 
selected the brightest, and best positioned to serve as reference stars. As 
stated above, ideally five or more reference
stars are desired for an FGS astrometry program. But the actual number of reference stars used
must be balanced vs. the time within a single {\it HST} orbit available for reference star measurement.
The fainter the object, the longer the time required for a position measurement. For U Gem and
SS Aur, only four reference stars were used due to the overall faintness of the reference stars and
the targets. For SS Cyg, a brighter target in a well populated reference field, five reference stars
were used. The positions, {\it VRI} photometry, spectral types, visual extinction, and spectroscopic
parallaxes of all thirteen reference stars employed in our program are listed in Table 1.

An observational sequence consists of slewing to the target field, acquisition of the target
DN, and then repeated measurement of the position of the target and each of the reference stars.
As described in the {\it Handbook}, the pointing of the {\it HST} exhibits a small drift throughout an orbit.
To account for this drift, one (or more) of the reference stars is chosen as a 
``drift check star''.
This star is measured more frequently than the other reference stars allowing this drift to be
modeled, and removed during data reduction. A typical observing sequence for SS Aur was: SS
Aur, Ref. \#2 (drift check star), Ref. \#12, Ref. \#21, Ref. \#9, SS Aur, Ref. \#2, Ref. \#12, Ref. \#21,
Ref. \#9, SS Aur, Ref. \#2, Ref. \#12, Ref. \#21, Ref. \#9, Ref. \#2. In the case of SS Aur, each star
was observed for 60 seconds. The acquisition of each object has an associated overhead, and the
combined exposure times and overheads for the observational sequence described above
consumed an entire 54 min {\it HST} ``orbit''. Using the {\it Handbook} guidelines, we estimated that two
such sequences were necessary at each epoch to achieve the program goal of parallaxes with a precision of $\leq$ 1 mas. 

During a single exposure time, a large number of independent samples of the 
target's position are collected. For objects with V = 14, the number of samples 
obtained in a 60 s exposure is $\approx$ 600. These samples are averaged to 
determine a single position in FGS coordinates.  At the end of a observational 
sequence, the {\it x} and {\it y} positions in FGS coordinates for all of the
targets are obtained. These positions then go thorough an extensive calibration process that
accounts for the OFAD, the pointing drift during the observation, and the differential velocity
aberration caused by the motion of the {\it HST} through space. The result is a set of positions for one
epoch of observation. To measure a parallax, observations at a minimum of two epochs are
necessary. To account for the proper motions of the target and reference stars, observations over
as long a timeline as possible are desired. To secure a set of observations approaching those
needed for a classical parallax measurement, we obtained data on three epochs. Each
observational epoch occurred at the season of the maximum parallax factor for the target DN.
Originally, these three epochs were separated by six months, but due to the difficulties with
NICMOS, and the subsequent changes in {\it HST} proposal priorities, our third epoch observations
were delayed by an additional six months. Thus, from start to finish, the observational program
spanned two years.

\section{PARALLAXES OF THE DWARF NOVAE}

After the observations were obtained, and processed, astrometric solutions 
were sought for each of the targets. As stated earlier, two observational 
sequences were obtained at each of
the three epochs. Thus, six independent sets of measurements were used in the astrometric
solution, performed by the Space Telescope Astrometry Team (STAT) at the University of Texas
(see Benedict et al. 1994). A master plate was constructed using a six parameter plate solution
that is simultaneously solved for translation, rotation, scale, and terms for independent scales on
the {\it x} and {\it y} axes. The solutions were robust for SS Cyg and U Gem, but less so for SS Aur.
During the first observational epoch for SS Aur, the FGS could not lock on to one of the
reference stars (Ref. \#8), apparently because it was much fainter than estimated in the {\it Guide Star
Catalogue}. Subsequently, for epochs two and three, we replaced this reference star with another
(Ref. \#21). Therefore, the solution for SS Aur was not as well constrained, and the resulting
precision was slightly poorer than found for the other two DN (see Harrison et al. 1999).

Relative parallaxes of the three DN were derived using two different astrometric
techniques: 1) assuming that the parallaxes and proper motions of the reference stars sum to zero,
and 2) that the reference frame is fixed (i.e., no motions are allowed in either parallax or proper
motion). Method \#1 is the technique preferred by the STAT, and is that used in normal ground
based parallax solutions. In all cases, however, both solutions were remarkably similar, producing
relative parallaxes that differed by only a few percent. This result suggests that the motions of
the reference stars were not significant. With the small number of reference stars used in our
program, however, a single reference star with a large parallax, or proper motion, could
dramatically effect the parallax of the program object. To evaluate each of our reference stars,
we treated them as the target and performed a solution to determine whether they exhibited large
parallaxes or proper motions. In no cases were large motions found, and the largest relative
parallax for a reference star was 2 mas. The final relative parallaxes for the DN are presented
in Table 2. The precision of these parallaxes, near $\pm$ 0.5 mas, exceeded the expected precision
by a factor of two. This added precision turned out to be especially important given the larger
than expected distances of the program DN.

     To convert the relative parallaxes to absolute parallaxes requires knowledge of the mean
parallax of the reference frame. To estimate this quantity two techniques are available. The first
is to determine the spectroscopic parallaxes for each reference star and average the results to
determine the reference frame parallax. The second technique relies on a model of the parallaxes
for stars at specific galactic coordinates, and within the magnitude ranges of the reference stars.
Both techniques were employed in this study. 

To determine the spectral types of the reference stars, moderate resolution 
(1.5 \AA /pix) optical spectroscopy of each object was obtained using the 
Double Beam Spectrograph on the 3.5 m telescope at Apache Point Observatory. 
These spectra were then compared to the digital atlas of MK standard spectral 
types by Jacoby, Hunter, and Christian (1984). The estimated spectral type for 
each reference star is listed in Table 1. To derive spectroscopic parallaxes,
optical {\it VRI} photometry was obtained to determine the visual magnitude and 
reddening for the reference stars. These data were acquired using the Clyde 
Tombaugh Observatory 16" Meade telescope located on the NMSU campus. This 
telescope is equipped with an SBIG ST-8 CCD camera with the standard Harris 
{\it UBVRI} filter set. All three DN fields were observed on
photometric nights along with Landolt standards. Standard techniques were used 
to derive the photometry of the reference stars listed in Table 1. A small 
number of {\it BVR} secondary standards, set up by Misselt (1996), are located within 
each DN field, and several of these happened to be reference stars in our 
program (identified in the notes column of Table 1). Differential
photometry was performed to check that our photometric solution reproduced the 
published values for these secondary standard stars. Using the spectral types 
and photometry, the visual extinction for each star was determined (very low in 
nearly all cases), and a spectroscopic parallax was estimated (final column of 
Table 1). The average values of these spectroscopic parallaxes are listed in 
column three of Table 2 as the observed reference frame parallax for each
DN field (we have assumed a 10\% error in the value of these parallaxes).

In order to compare our spectroscopic parallaxes with those predicted from the 
Yale model (van Altena, Lee, and Hoffleit 1995), we derived the reference frame 
parallaxes for all three DN fields. These results, listed in column four of 
Table 2, are in good agreement with the observed reference frame parallaxes 
computed from Table 1.

\section{RESULTS}
We derive the absolute parallaxes for each DN by correcting the FGS relative 
parallax by the observed reference frame parallax. The final results are listed 
in the penultimate column of Table 2. We convert these parallaxes to distances 
in the final column of Table 2. For SS Cyg, we derive a distance of 166.2 $\pm$ 
12.7 pc. This should be compared to published values of 30 $\pm$ 10 pc (Strand 
1948), 50 $\pm$ 15 pc (Kamper 1979), 76 pc (Warner 1987), $>$ 90 pc (Wade 1982),
95 pc (Bailey 1981), $>$ 95 pc (Berriman, Szkody, and Capps 1985), and 111 to 
143 pc (Kiplinger 1979). The Strand and Kamper results are astrometric 
parallaxes, while the other estimates used variations on the spectroscopic 
parallax of the secondary star. For U Gem we find a distance of 96.4 $\pm$ 4.6 
pc. Previous estimates were 76 pc (Wade 1979), 78 pc (Bailey 1981), 81 pc 
(Warner 1987), 100 $\pm$ 120 pc (van Maanen 1938), and 140 $\pm$ 70 pc 
(Berriman 1987). The van Maanen value is an astrometric parallax, while the 
other estimates used the photometric properties of the secondary star. For SS 
Aur we measured a distance of 200.0 $\pm$  25.7 pc. This should be compared
with a parallax-based distance of 100 $\pm$ 40 pc (Vasilevskis et al. 1975), 
spectroscopic parallaxes of $>$ 80 pc (Wade 1982) and $>$ 152 pc (Szkody and 
Mateo 1986), and a moving group parallax of 200 pc (Warner 1987).

It is clear that previous quotes for the astrometric parallaxes of these three 
DN were not significant. The parallax quoted with the greatest precision, that 
for SS Cyg by Kamper (1979), is clearly incorrect. The parallax of SS Cyg is 
much to small to have been detected using classical photographic astrometry. 
Additionally, the published parallax measurements for U Gem and SS Aur had such 
large error bars that they did not constitute actual detections of the parallax
for those objects. {\it Thus, we consider the three parallaxes listed in Table 2
to be the first true trigonometric parallaxes of any dwarf novae. }

The parallaxes derived here supply the first direct tests of the accuracy of 
secondary distance indicators in CVs. Except for Warner's estimate for the 
distance to SS Aur, and Kiplinger's estimate for SS Cyg, none of secondary 
methods provided distance estimates consistent with the astrometric parallaxes. 
By moving the spectroscopic parallax technique to the infrared, where 
contamination of the systemic luminosity by the white dwarf and accretion disk
is weaker, more precise spectroscopic parallaxes might be obtained. We examine 
the technique of infrared spectroscopic parallax for our program DN elsewhere 
(Harrison et al. 1999). Using new infrared data, and more recent calibrations of
the infrared luminosities of low-mass stars, we obtained accurate distance 
estimates for both SS Aur and U Gem. The infrared luminosity of SS Cyg, 
however, is apparently dominated by emission from the accretion disk. 

We have used the FGS on {\it HST} to measure the first trigonometric parallaxes 
of any dwarf novae. This modest program, consuming six {\it HST} orbits for each
object, has produced parallaxes with unrivaled precision. This increased 
precision over that foreseen when the proposal was submitted resulted from 
improved planning of the way the astrometric data is obtained with the
FGS, along with a greater understanding of how it should be to reduced. There 
is not a competing system in existence which can provide parallaxes of this 
precision on objects this faint in such a small amount of time. Clearly, the 
FGS on {\it HST} will not be supplanted as an astrometer until the development 
of the Space Interferometry Mission.

\acknowledgements
We would like to thank Denise Taylor and Ed Nelan for their help throughout our program.
Support for this work was provided by NASA through grant numbers GO-06538.01-95A and GO-
07492.01.96A from the Space Telescope Science Institute, which is operated by the Association
of Universities for Research in Astronomy under NASA contract NAS 5-26555.

\clearpage
\ptlandscape
\begin{deluxetable}{lccccccccc}
\scriptsize
\tablewidth{0pt}
\tablecaption{Reference Star Parameters}
\tablehead{
\colhead{Ref. Star} & \colhead{$\alpha_{\rm 2000}$} & \colhead{$\delta_{\rm 2000}$} & \colhead{V} & \colhead{V $-$ R} & \colhead{V $-$ I} & \colhead{Sp. Type}
& \colhead{A$_{\rm V}$}&\colhead{$\pi$ (mas)} & \colhead{Notes}}
\startdata
SS Aur \#2&06$^{\rm h}$:13$^{\rm m}$:29$^{\rm s}$.4&+47$^{\circ}$:44':26.5"&14.64&0.46&0.81&K0V&0.0&1.77&= Misselt \#5\nl
SS Aur \#9&06:13:44.4	&+47:44:53.8&12.54&0.31&0.59&F8I/II&1.0&$<$ 0.17&\nodata
\nl

SS Aur \#12&06:13:09.8&+47:42:56.0&16.10&0.46&1.03&K0V&0.0&0.90&\nodata \nl

SS Aur \#21&06:13:46.5&+47:43:22.9&14.17&0.51&0.89&K0V&0.0&2.20&\nodata \nl

U Gem \#2&07:55:07.1&+21:59:19.2&14.77&0.40&0.77&K0V&0.0&1.67&= Misselt \#4\nl

U Gem \#4&07:55:04.6&+22:01:53.1&13.95&0.31&0.61&F4V&0.3&0.65&= Misselt \#1\nl

U Gem \#8&07:55:23.4&+21:59:57.1&12.01&0.34&0.56&G0V&0.0&3.00&\nodata \nl

U Gem \#9&07:55:24.1&+21:58:39.8&14.20&0.33&0.61&G0V&0.0&1.10&\nodata \nl

SS Cyg \#3&21:42:43.4&+43:34:23.6&13.36&0.27&0.39&F0V&0.18&0.80&= Misselt \#6\nl

SS Cyg \#6&21:42:33.2&+43:34:02.6&12.05&1.06&2.20&K4III&1.31&0.98&= Misselt \#1\nl

SS Cyg \#7&21:42:27.1&+43:33:44.7&10.80&0.32&0.58&F8V&0.0&4.37&\nodata \nl

SS Cyg \#12&21:42:35.5&+43:35:44.0&15.11&0.76&1.59&G5V&0.18&1.08&\nodata \nl

SS Cyg \#14&21:42:58.7&+43:35:18.0&12.92&0.40&0.70&G0V&0.18&2.15&\nodata \nl

\enddata
\end{deluxetable}

\ptlandscape
\begin{deluxetable}{lccccc}
\scriptsize
\tablewidth{0pt}
\tablecaption{Object and Reference Frame Parallaxes, and the Distances to the
Program Dwarf Novae}
\tablehead{
\colhead{Dwarf Nova}&\colhead{Relative Parallax}&\colhead{Observed Reference}&\colhead{Model Reference}&\colhead{Parallax (mas)}&\colhead{Distance (pc)}\\
\colhead{}&\colhead{(mas)}&\colhead{Parallax (mas)}&\colhead{Parallax (mas)}&
\colhead{}&\colhead{}}
\startdata

SS Aur&3.74 $\pm$ 0.63&1.26 $\pm$ 0.13&1.2&5.00 $\pm$ 0.64&200.0 $\pm$ 25.7 \nl
U Gem&8.77 $\pm$ 0.47&1.60 $\pm$ 0.16&1.4&10.37 $\pm$ 0.50&96.4 $\pm$ 4.6 \nl
SS Cyg&4.14 $\pm$ 0.42&1.88 $\pm$ 0.19&1.7&6.02 $\pm$ 0.46&166.2 $\pm$ 12.7 \nl
\enddata
\end{deluxetable}

\end{document}